# On Demand Cell Sectoring Based Fractional Frequency Reuse in Wireless Networks

Shakil Ahmed, Mohammad Arif Hossain, and Mostafa Zaman Chowdhury
Department of Electrical and Electronic Engineering
Khulna University of Engineering & Technology, Khulna-9203, Bangladesh
E-mail: shakileee076@gmail.com, dihan.kuet@gmail.com, mzceee@yahoo.com

*Abstract*—In this paper, a dynamic channel assigning along with dynamic cell sectoring model has been proposed that focuses on the Fractional Frequency Reuse (FFR) not only for interference mitigation but also for enhancement of overall system capacity in wireless networks. We partition the cells in a cluster into two part named centre user part (CUP) and edge user part (EUP). Instead of huge traffic, there may be unoccupied channels in the EUPs of the cells. These unoccupied channels of the EUPs can assist the excessive number of users if these channels are assigned with proper interference management. If the number of traffic of a cell surpasses the number of channels of the EUP, then the cell assigns the channels from the EUP of other cells in the cluster. To alleviate the interference, we propose a dynamic cell sectoring scheme. The scheme sectors the EUP of the cell which assigns channels that the assigned channels are provided to the sectored part where these channels receive negligible interference. The performance analysis illustrates reduced call blocking probability as well as better signal to interference plus noise ratio (SINR) without sacrificing bandwidth utilization. Besides, the proposed model ensures lower outage probability.

I. INTRODUCTION

Limited radio resource is inadequate to maintain different wireless services with proper Quality of Service (QoS). In this case, an effective allocation of resources can be greatly helpful to provide the subscribers with desired QoS. It also improves the system capacity in wireless networks. Fractional Frequency Reuse (FFR) is one of the effective ways for resource allocation in traffic congested network [1]. However, frequency reusing technique creates heavy inter-cell interference problem and the users at the cell edge might feel degradation in connection quality. There creates an emergence to mitigate interference [2].

Soft Frequency Reuse (SFR) is studied for FFR and interference mitigation in earlier work [2]-[6]. Though the performance with the traditional SFR scheme might be better, the resource may keep unutilized in case of lower traffic condition. Moreover, the edge part of any cell under this scheme gets more area compared to the centre. Consequently, the users at the edge need more bandwidth to get connected in the network. Keeping these pitfalls under consideration, we proposed dynamic FFR scheme along with dynamic sectoring model to mitigate interference. We develop the scheme ensuring maximum utilization of unoccupied channels with interference declination. There may be unoccupied channels in the cells instead of huge traffic. A cell having the number of users greater than the total number of channels can take the advantage by assigning channels from adjacent cells which have those unoccupied channels [7]. According to our proposed scheme, when cells have less traffic at the cell edge, these can lend the channels of the edge to the cell which has highest traffic intensity. Subsequently, the unutilized channels become activated and better QoS can be ensured for the users who may become the victim of poor QoS. However, the assigning of the reused frequency causes huge interference as the frequencies are available in the nearest adjacent cells. To reduce the interference, a dynamic sectoring model has been proposed. As the channel assigning process is dynamic, the interference mitigation process is also dynamic. It has also the advantage over static sectoring scheme as the sectoring of cell causes frequent handover. Though the sectoring has some disadvantages but it can reduce interference. Thus the proposed model surpasses the limitations because bandwidth utilization is one pivotal point for a successful wireless communication.

The rest of the paper is organized as follows: Section II shows the proposed scheme with proper illustration. The queuing analysis for the proposed model is shown in Section III. Section IV contains the outage probability analysis while the simulation results of the proposed scheme are placed in section V. Finally, conclusions are drawn in Section VI.

II. DYNAMIC CELL SECTORING

Imminent wireless networks need to connect maximum number of users with reduced overall call blocking probability.

*A. System model*

In our proposed model we assume a cluster of seven cells with reused frequency band. The frequency band is divided into six segments that are defined as $F_1$, $F_2$, $F_3$, $F_4$, $F_5$, and $F_6$ for a cell. Each cell has total $S$ number of original channels. According to the proposed model, each cell of the cluster is bifurcated. The inner and outer parts are called centre user part (CUP) and edge user part (EUP), respectively. Figure 1 shows the initial frequency band allocation. Cell 1, cell 3, and cell 5 are provided with frequency band $C$ for CUP and $F_5$, $F_6$ for EUP. Frequency band $C$ comprises the band from $F_1$ to $F_4$. Similarly frequency bands $A$ ($F_3$-$F_6$) is allotted for the CUP of cell 7 whereas $F_1$ and $F_2$ are for EUP of the cell. Finally band

B ($F_1$, $F_2$ and $F_5$, $F_6$) is for the CUP of cell 2, cell 4, and cell 6. $F_3$ and $F_4$ are provided to EUP of cell 2, cell 4, and cell 6. CUP of cell 7 has the frequency bands $F_3$ to $F_6$. We consider cell 7 as reference cell that has the highest traffic. So reduced call blocking probability and insignificant interference are mandatory for cell 7 in order to achieve better QoS. We assume two categories of cells except cell 7 according to their frequency band. First category has cell 1, cell 3, and cell 5 while second category has cell 2, cell 4, and cell 6. Cell 3 is assumed having lower traffic intensity among the cells of first category whereas cell 2 is for second category. Let the cell having highest number of unused channels of first category i.e. cell 3 is defined as $\alpha_{h,uc}$ and cell 2 of second category is $\beta_{h,uc}$

### B. Dynamic Channel Assigning Architecture

In our proposed model we assign the channels from the EUP of the adjacent cells to EUP of cell 7 dynamically. The CUPs of the cells are not our concern as the interference problem in the CUPs is negligible. $S_{need}$ is the required number of channels. So $\alpha_{h,uc}$ of first category is searched i.e. cell 3. If the unoccupied channels $S_{\alpha h,uc}$ of cell 3 is larger compare to the $S_{need}$ of cell 7, the cell assign $S_{need}$. If $S_{need}$ is greater, then $S_{\alpha h,uc}$ is assigned. So total number of channels in cell 7 is $S+ S_{\alpha h,uc}$. Thus the total number of needed channels are not fulfilled. So $S_{need(additional)}$ number of channels are required. Now the cells of second category is searched, cell 2 i.e. $\beta_{h,uc}$ is found with highest number of unused channels. If the unused channels $S_{\beta h,uc}$ of cell 3 is higher than $S_{need(additional)}$, after assigning the total number of channels in cell 7 becomes $S+S_{\alpha h,uc}+S_{need(additional)}$. If it is not, then total channels in the reference cell is $S+S_{\alpha h,uc}+ S_{\beta h,uc}$. This assigning procedure is shown in the block diagram of Fig 2.

### C. Dynamic Cell Sectoring for Interference Management

To reduce interference, cell sectoring is introduced in cell 7. The cell is sectored by 120 degree. The sectored EUP of cell 7 has two parts, named X and Y. It is shown in Fig 2 that frequency band $F_3'$ and $F_5'$ of EUP of cell 2 and cell 3 can be assigned by cell 7 if the band remains unused. Similarly, the frequency band $F_4'$ and $F_6'$ of EUP of cell 2 and cell 3 can be assigned by cell 7 if frequency band remains unused. Here $0 \leq F_3' \leq F_3$, $0 \leq F_4' \leq F_4$, $0 \leq F_5' \leq F_5$ and $0 \leq F_6' \leq F_6$.

### III. QUEUING ANALYSIS

The call arriving process of the proposed scheme is assumed to be Poisson and the queuing process can be modeled as $M/M/K/K$. The Markov chain of the initial condition before assigning channels and for the proposed scheme are shown in Fig. 4 and Fig. 5, respectively. $\lambda_7$, $\lambda_2$ and $\lambda_3$ indicate the call arrival rates for cell 7, cell 2, and cell 3, respectively. $\mu_7$, $\mu_2$, and $\mu_3$ indicate the channel release rate for cell 7, cell 2, and cell 3, respectively. Besides, $S_7$ and $S_7'$ represent the total number of channels in cell 7 (reference cell) before and after assigning channels, respectively where $S_7'>S_7$. Similarly, $S_2$ and $S_3$ represent the total number of channels in cell 2 and cell 3 before channel assigning process, respectively, whereas $S_2'$ and $S_3'$ represent the total number of channels in cell 2 and cell 3 after channel assigning process, respectively where $S_2' \leq S_2$ and $S_3' \leq S_3$. According to our assumption, the number of cells in a cluster be $M$ and $S_m$ and $S_m'$ are the maximum number of calls in a cell $m$ before and after channel assigning, respectively. The call arrival rate and channel release rate for cell $m$ are denoted by $\lambda_m$ and $\mu_m$, respectively, whereas $P_m(k)$ be the steady state probability of the system in state $k$ for cell $m$. The call blocking probability $P_{Bm}$ and overall call blocking probability $P_T$ of the system can be expressed as

$$P_{Bm} = \frac{\lambda_m^{S_m}}{S_m! \mu_m^{S_m}} P_m(0) \qquad (1)$$

$$P_T = 1 - \frac{\sum_{m=1}^{M} \lambda_m (1-P_{Bm})}{M \times \sum_{m=1}^{M} S_m} \qquad (2)$$

### IV. OUTAGE PROBABILITY ANALYSIS

The received SINR level for the users can be expressed as

$$SINR = \frac{S_0}{\sum_{l_1=1}^{M_1} I_1(l_1) + \sum_{l_2=1}^{M_2} I_2(l_2) + ... + \sum_{l_u=1}^{M_u} I_u(l_u) + ... + \sum_{l_U=1}^{M_U} I_U(l_U)} \qquad (3)$$

where $S_0$ is the received signal power from the BS, $u$ denotes a tier among the interfering tiers. $l_u$ and $M_u$ represent a cell among the interfering cells of $u$-th tier and maximum of the interfering cells in $u$-th tier, respectively. Besides, $I_U(l_u)$ refers the received interfering power from $l_u$-th cell of $u$-th tier.

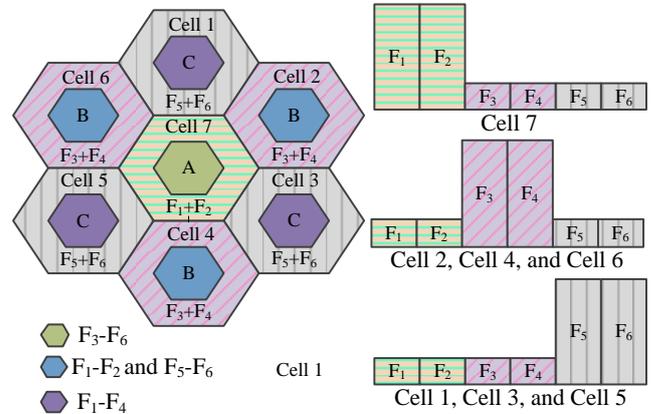

**Fig. 1.** Initial frequency band allocation.

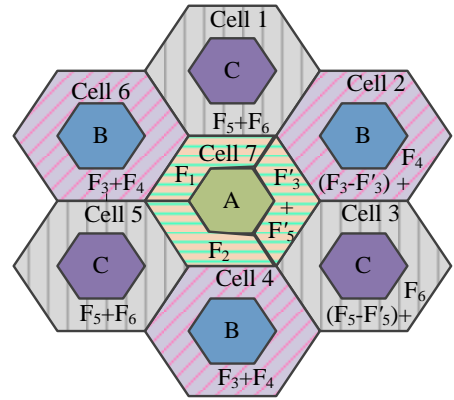

**Fig. 2.** Proposed channel assigning architecture along with dynamic cell sectoring assigning partial bandwidth spectrum.

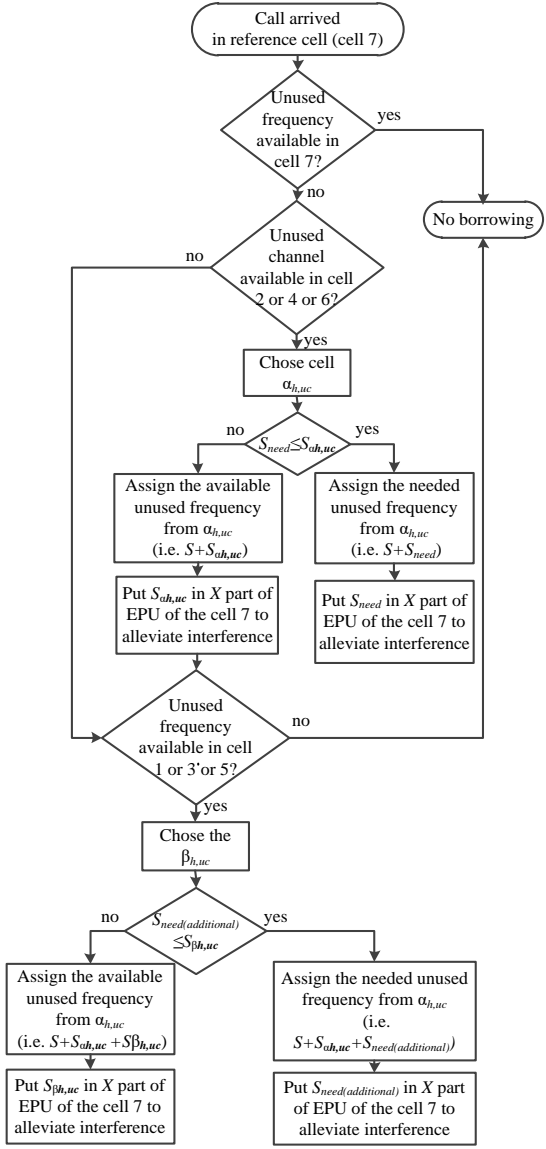

**Fig. 3.** Block diagram of dynamic channel assigning and interference management.

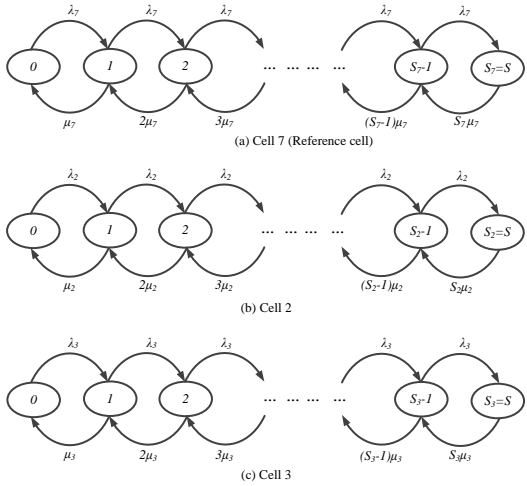

**Fig. 4.** Markov chain before assigning channels.

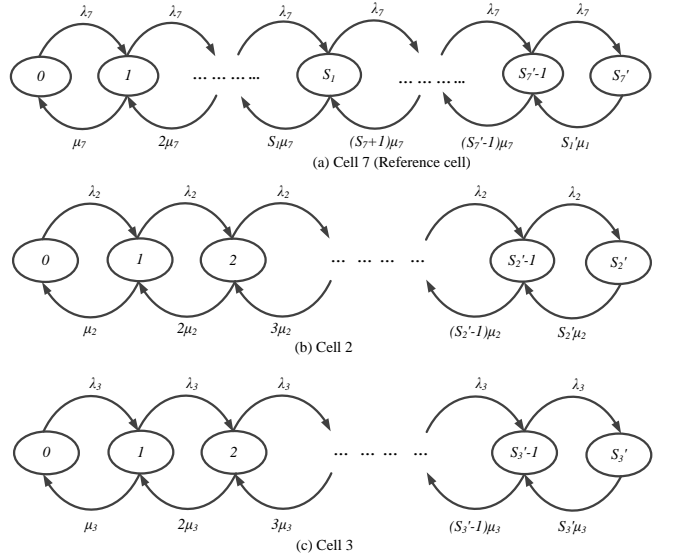

**Fig. 5.** Markov chain after assigning channels.

The outage probability [8] can be expressed as

$$P_{out} = \Pr(SINR < \gamma) \qquad (4)$$

where $\gamma$ stands for the threshold value (9 dB) of SINR below which the connection of the user cannot be sustained.

## V. SIMULATION RESULTS

We compare our proposed scheme with conventional scheme with frequency reuse factor 3. Conventional model is defined where the cell is bifurcated but not sectored on demand basis. We assume a cluster of seven cells and the call arrival rate ratio of the 7 cell of the cluster as 7:1:2:4:5:5:6. The summation of the corresponding call arrival rate of the cell in the cluster is defined as total call arrival rate. The fundamental parameters used for the performance analysis of our scheme are depicted in Table I. Figure 6 expresses the reduced call blocking probability of the reference cell of the proposed scheme even if the cell has higher traffic intensity. It proves the supremacy of the proposed scheme compared to conventional model as it shrinks the call blocking probability of the system. The efficient cellular communication requires the maximum bandwidth utilization. Figure 7 shows the comparison of the bandwidth utilization of the proposed scheme and the conventional scheme. Our scheme shows better performance.

The comparison of SINR levels of the proposed scheme and the conventional scheme is demonstrated in Fig 8. The proposed model shows better SINR level. As the cell, having higher traffic intensity, are sectored and the unused channels of the adjacent cells are provided to the EUP of the sectored cell, this receives strong signal from the BS of the sectored cell and trifling interference from the BS of the adjacent cells. Thus the proposed model proves the novelty in wireless networks. Figure 9 represents the comparison of the outage probability of the proposed scheme and conventional model. The outage probability of the proposed scheme is expressively smaller compared to conventional model.

TABLE I: Parameter values used in the performance analysis.

| Parameter | Value |
| --- | --- |
| Number of original channels in each cell | 120 |
| Threshold value of channel assigning in a cell | 80 |
| Transmitted signal power by the BS | 1.50 kw |
| Height of the BS | 100 m |
| Average channel holding time | 90 sec |
| Cell radius | 1Km |

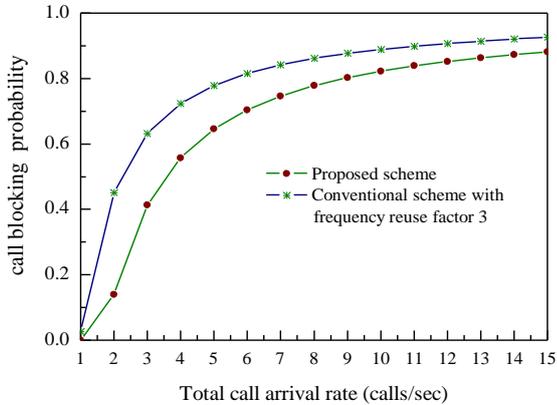

**Fig. 6.** Comparison of call blocking probability of reference cell.

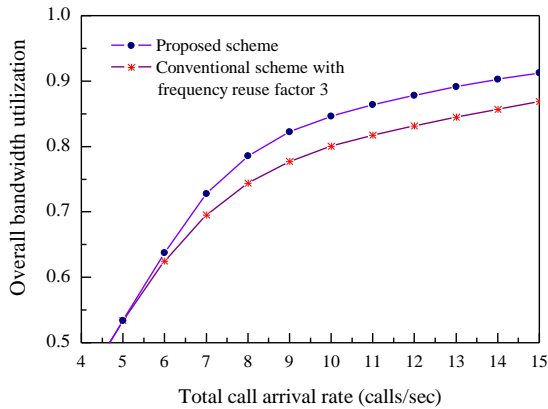

**Fig. 7.** Comparison of bandwidth utilization.

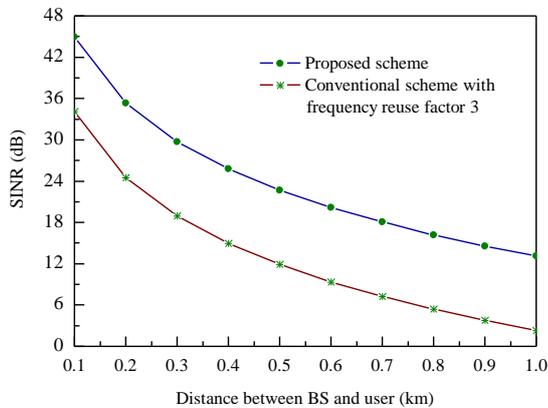

**Fig. 8.** Comparison of SINR levels.

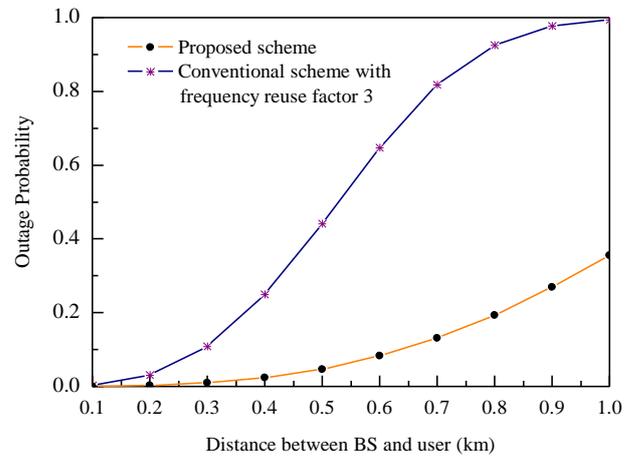

**Fig. 9.** Comparison of outage probability.

## VI. CONCLUDING NOTES

The proposed model ensures the utilization of maximum bandwidth and reduced overall call blocking probability. More importantly introduction to dynamic cell sectoring mitigates the interference problem which is the most prominent feature of our proposed model. Using of reused frequency bands and put them into the sectored part of the reference cell so that minimum interference occurs, are the important findings of our proposal. As the performances of our scheme are highly efficient, it can be a better choice for the operators in future.